\documentclass[12pt,prc,onecolumn,final]{revtex4}
\usepackage{graphicx}
\usepackage{amssymb}
\usepackage{epstopdf}
\def\het{\ifmmode {}^3{\rm He} \else $^3$He\fi}
\def\trit{\ifmmode {}^3{\rm H} \else $^3$H\fi}
\def\ee{\ifmmode (e,e') \else $(e,e')$\fi}
\def\eep{\ifmmode (e,e'p) \else $(e,e'p)$\fi}
\def\een{\ifmmode (e,e'n) \else $(e,e'n)$\fi}
\def\deep{\ifmmode d(e,e'p) \else $d(e,e'p)$\fi}

\begin{document}

\title{ Proton and Neutron Momentum Distributions in $A=3$ Asymmetric
  Nuclei \\
A Hall A Collaboration Experiment \\
Proposal PR12-13-012 to  Jefferson Lab PAC 42, July 2014}

% \newcommand*{\ODU}{Old Dominion University, Norfolk, Virginia 23529}
% \newcommand*{\ODUindex}{1}
% \affiliation{\ODU}
% \newcommand*{\JLAB}{Thomas Jefferson National Accelerator Facility, Newport News, Virginia 23606}
% \newcommand*{\JLABindex}{2}
% \affiliation{\JLAB}
% \newcommand*{\TAU}{Tel Aviv University, Tel Aviv, Israel}
% \newcommand*{\TAUindex}{2}
% \affiliation{\TAU}

\author{\baselineskip 13 pt
  C. Hyde, S.E. Kuhn and L.B. Weinstein (co-spokesperson) \\
 Old Dominion University, Norfolk VA \\ [0.2cm]
 M. Braverman, E. Cohen, O. Hen (co-spokesperson), I. Korover, J. Lichtenstadt, E. Piasetzky, and  I. Yaron \\
 Tel-Aviv University, Tel Aviv, Israel \\ [0.2cm]
 W. Boeglin (co-spokesperson), P. Markowitz and M. Sargsian\\
 Florida International University, Miami, FL\\ [0.2cm]
 W. Bertozzi, S. Gilad (co-spokesperson), and V. Sulkosky\\
 Massachusetts Institute of Technology, Cambridge, MA \\ [0.2cm]
D.W. Higinbotham, C. Keppel, P. Solvignon and S.A. Wood \\
Thomas Jefferson National Accelerator Facility, Newport News, VA \\
[0.2cm]
 Guy Ron \\
 Hebrew University, Jerusalem, Israel \\ [0.2cm]
R. Gilman \\
Rutgers University, New Brunswick, NJ \\ [0.2 cm]
 J.W. Watson \\
 Kent State University, Kent, OH \\ [0.2cm]
 A. Beck and S. Maytal-Beck \\
 Nuclear Research Center Negev, Beer-Sheva, Israel \\ [0.2cm]
J. Beri\v{c}i\v{c}, M. Mihovilovi\v{c}, S. \v{S}irca, and
S. \v{S}tajner \\
Jo\v{z}ef Stefan Institute, Ljubljana, Slovenia \\ [0.2 cm]
D. Keller \\
University of Virginia, Charlottesville, VA \\ [0.2 cm]
Vincenzo Bellini, Maria Concetta Sutera and Francesco Mammoliti \\
INFN/CT and University of Catania, Catania, Italy \\ [0.2 cm]
J. Annand, D. Hamilton, and D. Ireland \\
University of Glasgow, Glasgow, United Kingdom \\ [0.2cm]
A. Sarty \\
St. Mary's University, Halifax, Nova Scotia, Canada \\ [0.2cm]
L. Kaptari \\
Bogoliubov Lab. Theor. Phys., 141980, JINR, Dubna,  Russia \\ [0.2 cm]
C. Ciofi degli Atti \\
INFN Perugia, Perugia, Italy \\ [0.2 cm]
 }
\date{\today}                                           % Activate to display a given date or no date

%\begin{titlepage}

% \begin{center}
% \vspace{-0.3cm}
% {\Large Nucleon Momentum Distributions in Asymmetric Nuclei \\
% A Comparison of $^3$He$(e,e'p)$ and $^3$H$(e,e'p)$} \\[0.4cm] 
% A proposal to Jefferson Lab PAC 40, June 2013 \\ [0.4cm]

%  L.B. Weinstein (co-PI) \\
%  Old Dominion University, Norfolk VA 23529\\ [0.2cm]
%  M. Braverman, O. Hen, I. Korover, J. Lichtenstadt, E. Piasetzky (co-PI), and  I. Yaron \\
%  Tel-Aviv University, Tel Aviv, Israel \\ [0.2cm]
%  Werner Boeglin (co-PI) and Misak Sargsian\\
%  Florida International University, Miami\\ [0.2cm]
%  W. Bertozzi, Shalev Gilad (co-PI), and Vincent Sulkosky\\
%  Massachusetts Institute of Technology, Boston \\ [0.2cm]
%  Guy Ron \\
%  Hebrew University, Jerusalem, Israel \\ [0.2cm]
%  J.W. Watson \\
%  Kent State University, Kent, OH \\ [0.2cm]
%  A. Beck and S. Maytal-Beck \\
%  Nuclear Research Center Negev, Beer-Sheva, Israel \\ [0.2cm]
% D.W. Higinbotham, S. Stepanyan, B. Sawatzky, and S. A. Wood \\
% Thomas Jefferson National Accelerator Facility, Newport News, VA \\ [0.2cm]

% \end{center}
%\end{titlepage}

\begin{abstract}
  \newpage \baselineskip 15pt In {\it non-interacting} Fermi systems
  with imbalanced number of two different Fermions, the average
  momentum per fermion is higher for the majority. Adding a strong
  short-range interaction between different fermions may invert the
  momentum sharing of the two components, making the minority move
  faster on average than the majority. This feature is due to the high
  momentum distribution being dominated by short distance pairs of
  different type Fermions. It is a common behavior that applies to
  systems ranging from ultra-cold atoms at neV energies to nucleons
  with MeV energies.  In nuclei the nucleon-nucleon tensor force makes
  the neutron-proton short range correlated pair ($np$-SRC) the
  dominant component contributing to the high momentum of nucleons.
  In light neutron-rich nuclei such as \het, the average momentum of a proton
  should be higher than that of a neutron. In $^3$He the average
  momentum of the neutron should exceed that of the protons.
 
  We propose to verify the above prediction by measuring both the
  majority and minority nucleon momentum distributions of asymmetric
  $A=3$ nuclei. We will do this by measuring the quasielastic
  \trit\eep{} and \het\eep{} reactions at $Q^2\approx 2$ (GeV/c)$^2$
  and $x_B=1.2$ for missing momenta up to 450 MeV/c using the MARATHON
  target and the two HRS spectrometers in Hall A in kinematics which
  minimize the effects of Final State Interactions.  Because the
  MARATHON target is limited to low luminosity, dead times and other
  rate effects will be negligible.  

  We plan to measure (a) the proton momentum distributions of both
  \het{} and \trit{} in order to constrain detailed calculations of
  the $A=3$ system, (b) the ratio of \het\eep/\trit\eep{} cross
  sections where the residual FSI will mostly cancel, and (c) the
  average kinetic energy of protons in the two nuclei as a function of
  the maximum measured momentum.  It is crucial to measure these in
  $A=3$ nuclei since these are the simplest asymmetric mirror nuclei
  and the only ones for which precisions calculations of the \eep{}
  reaction exist.

This proposal was deferred by PAC 40.  In this  resubmission, we have
reduced the requested beam time from 30 to 10 days by focussing on
\het{} and \trit, omitting the deuterium target and by reducing
the beam time at large missing momentum.  Because the MARATHON target
is scheduled to be installed in Fall 2015, this is the last
opportunity to approve any experiment to measure \trit\eep{} at JLab.

  We request 1 day of beam time for calibration and 9 days for the
  measurements of the \trit\eep{} and \het\eep{} reactions. 
  This will be the first \eep{} measurement of \trit{} and the only one
  on \het{} in reduced-FSI kinematics.  
\end{abstract}

\maketitle

\thispagestyle{empty}
%
%\pagenumbering{roman}
\setcounter{page}{1}
\newpage

\section{Introduction and Motivation}

\subsection{Common features of balanced (symmetric) 
and imbalanced (asymmetric) interacting two component Fermi systems}

Many-body systems of two types of interacting fermions play an
important role in nuclear, astro, atomic, and solid-state
physics. Particularly intriguing are systems that include a
short-range interaction that is strong between different fermions and
weak between fermions of the same type.

Recent theoretical advances show that even though the underlying
interaction between the fermions can be very different, these systems
present several common features
\cite{Tan08a,Tan08b,Tan08c,Braaten12}. These features include the
existence of a high-momentum tail ($k > k_F$, where $k$ is the fermion
momentum and $k_F$ the Fermi momentum) that scales as $C/k^4$ and is
dominated by short-range correlated (SRC) pairs of different
fermions. The SRC pairs have large relative momentum between the
fermions and small center-of-mass (CM) momentum, where large and small
is relative to the Fermi momentum of the system. The scale factor,
$C$, is known as Tan's contact and fully controls the thermodynamics
of the system \cite{Braaten12}.

The application to the nuclear case is more complicated because
the range of the nuclear short-range tensor interaction is not much much smaller than the
inter-nucleon separation and because it is a tensor interaction.  This
leads to a momentum density $n(k)$ at high momentum ($k > 1.5 k_F$) proportional to $V^2/k^4$
where $V$ is the momentum-dependent strength of the tensor interaction.

However, recent experimental studies of balanced (symmetric) interacting Fermi
systems, with an equal number of fermions of two kinds, showed that
the SRC pairs are predominantly unlike-fermion pairs
\cite{Stewart10,subedi08,piasetzky06,egiyan06,fomin12}. These
experiments were done using very different Fermi systems: protons and
neutrons in atomic nuclei \cite{egiyan06,fomin12} and two-spin state
ultra-cold atomic gasses \cite{Stewart10}, which span more than 15
orders of magnitude in Fermi energy from  MeV to neV), and exhibit
different short-range interactions (a strong tensor interaction in the
nuclear systems \cite{subedi08,schiavilla07}, and a tunable Feshbach
resonance in the atomic system \cite{Stewart10}).  

\begin{figure}[htpb]
\begin{minipage}[b]{3. in}
\begin{center}
\includegraphics[width=3.3in]{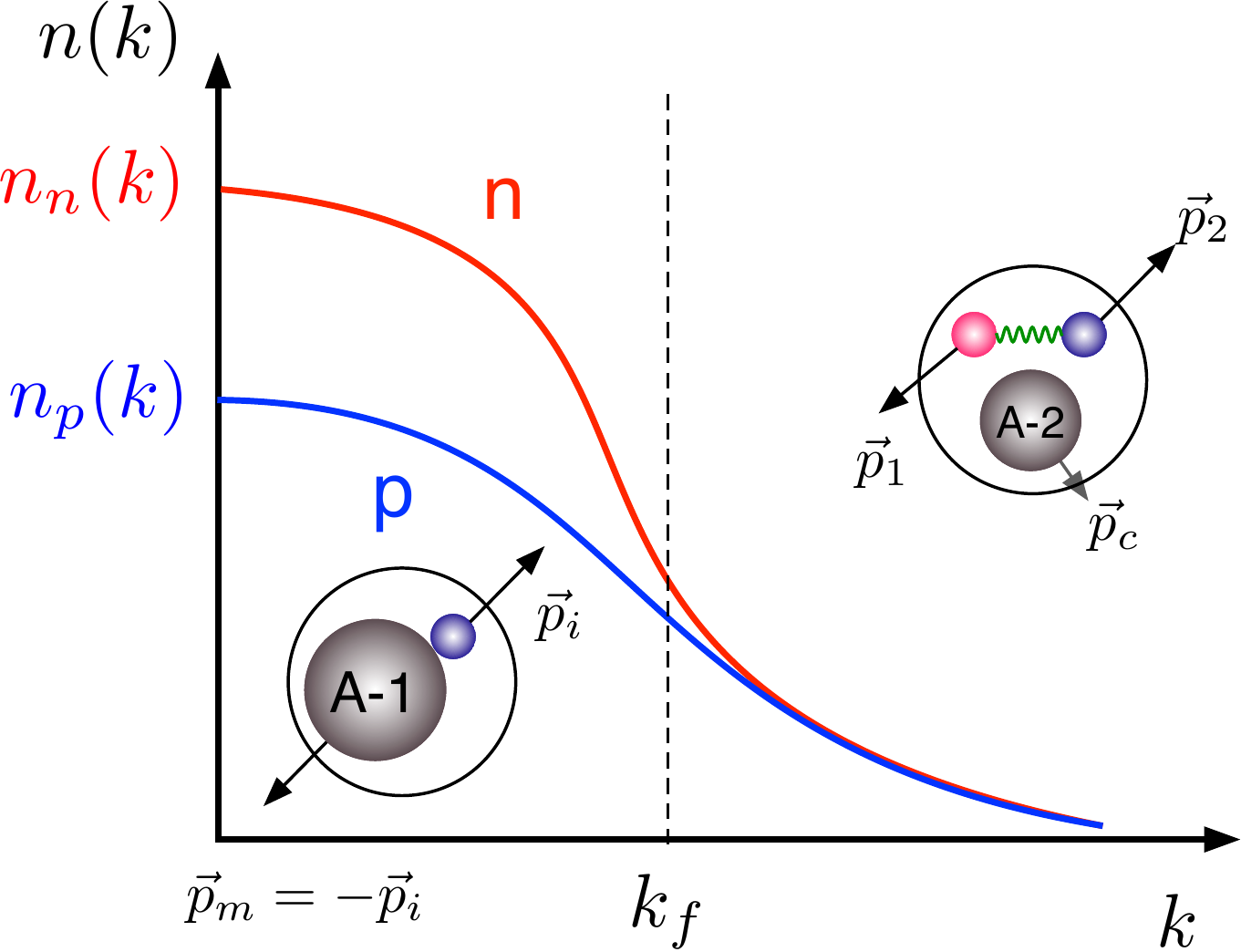}
\end{center}
     \end{minipage}\hfill
\vspace{-1.2in}
     \begin{minipage}[t]{3in}
\vspace{-2in}
\caption[]{\label{fig:MomDistCartoon1} \baselineskip 13pt Illustration
  of the momentum distribution of protons and neutrons in asymmetric
  nuclei. $n_n(k)$ and $n_p(k)$ are the neutron and proton momentum
  densities, respectively. 
  The insets show the single nucleon in a mean field and the 2N-SRC
  pairs that dominate below and above $k_F$, respectively.}
\end{minipage}
%\end{center}
\end{figure}

The nuclear studies showed that in the symmetric $^4$He and $^{12}$C
nuclei, every proton with momentum $300 < k<600$ MeV/c has a
correlated partner nucleon, with neutron-proton ($np$) pairs
outnumbering proton-proton ($pp$) and, by inference, neutron-neutron
($nn$) pairs by a factor of $\approx 20$
\cite{subedi08,piasetzky06,korover14}.  A recent study showed that in 
asymmetric heavy nuclei with unequal numbers of the
different fermions, high-momentum protons still
disproportionately belong to $np$ pairs \cite{hen14}.  This suggested
a new feature: an inversion of the momentum sharing between the
minority and majority components. This inversion is due to the
short-range interaction, which populates the high-momentum tail with
equal amounts of majority and minority fermions, thereby leaving a
larger proportion of majority fermions to occupy low momentum states
$(k<k_F)$ (see Fig.{} \ref{fig:MomDistCartoon1}).

For light asymmetric nuclei the simple concept described above can be
illustrated by detailed microscopic calculations.

\vspace {1.2in}

\subsection{Energy Sharing in Light Nuclei ($A<12$)}

Detailed Variational Monte Carlo  (VMC) calculations of the single-nucleon momentum
distributions for a variety of symmetric and asymmetric light nuclei
$(2\le  A \le 12)$ are now available \cite{wiringa14}. By integrating
these single-nucleon momentum distributions one can obtain the average
proton and neutron kinetic energies for different nuclei:
\begin{equation}
\langle T_p\rangle = \frac{\int_0^\infty n_p(k)(\sqrt{m_p^2+k^2}-m_p)d^3k}{\int_0^\infty
  n_p(k)d^3k}
\label{eq:Tn}
\end{equation}
where $n_p(k)$ is the proton momentum distribution (and simlarly for neutrons).  

The average kinetic energy of the minority nucleons is larger than
that of the majority nucleons in asymmetric nuclei and this effect
increases with the nuclear asymmetry, see Table \ref{tab:TN}. This
non-trivial result can be naturally explained by the dominance of
neutron-proton pairs in the high momentum tail of the nuclear momentum
distribution.

\begin{table}[htb]
\begin{center}
{\baselineskip 13pt
\begin{tabular}{|c|c|c|c|c|} \hline
Nucleus & Asymmetry & $\langle T_p \rangle$ & $\langle T_n \rangle$ &
$\langle T_p \rangle / \langle T_n \rangle$ \\
 & $(N-Z)/A$ &&& \\ \hline
$^8$He & 0.50 & 30.13 & 18.60 & 1.62 \\ \hline
$^6$He & 0.33 & 27.66 & 19.60 & 1.41 \\ \hline
$^9$Li & 0.33 & 31.39 & 24.91 & 1.26 \\ \hline
{\bf $^3$He} & $-0.33$ & 14.71 & 19.35 & {\bf 0.76} \\ \hline
{\bf $^3$H} & 0.33 & 19.61 & 14.96 & {\bf 1.31} \\ \hline
$^8$Li & 0.25 & 28.95 & 23.98 & 1.21 \\ \hline
\end{tabular}
}
\end{center}
\vspace{-13 pt}
\caption{\baselineskip 13pt The proton and neutron average kinetic
  energies as extracted  from ab-initio VMC single-nucleon momentum 
  distribution calculations \cite{wiringa14}. The average kinetic energy
  of the minority nucleons is larger than that of the majority
  nucleons.  This difference increases with the nuclear asymmetry \cite{sargsian14}. 
}
\label{tab:TN}
\end{table}

We propose here to measure the momentum distribution of protons in the
most asymmetric stable nuclei, \het{} and \trit, and demonstrate that
the minority (protons in \trit) have average kinetic energy higher
than the majority (protons in \het). The $A=3$ nuclear system is a unique laboratory for
such a study as discussed below.

\subsection{Why $^3$He and $^3$H?}

The $A=3$ system is the lightest and simplest asymmetric nuclear
system. We can use the fact that that \het{} and \trit{} are mirror
nuclei to measure the difference between the momenta carried by
majority ($n$ in \trit{} and $p$ in \het) and minority ($p$ in \trit{}
and $n$ in \het) fermions by measuring \het\eep{} and \trit\eep.  By
isospin symmetry, we expect the proton momentum distribution in
\trit{} to equal the neutron momentum distribution in \het{} and vice
versa. This avoids the difficulties and inaccuracies due to detecting
neutrons.

Nucleon ground state momentum
distributions cannot be directly measured;
we can only measure missing-momentum distributions which represent a
convolution of ground state momentum distributions and final state
interaction effects.  However, by measuring the relatively simple $A=3$ nuclei
we can (a) choose kinematics that 
minimize the effects of FSI (see Section \ref{sec:FSI}) and (b) calculate the
relatively small correction due to FSI (see section \ref{sec:FSI}).

%\subsection{Majority and minority fermions in  $A=3$ nuclei}
There are no measurements of neutron momentum distributions in \het{}
or of neutron or proton momentum distributions in \trit. 

We therefore propose to measure the majority and minority momentum
distributions and their ratio with high accuracy by measuring the quasielastic
knock-out of protons from \het{} and \trit.  Since there are many more
calculations of $^3$He, we will show plots of
neutron and proton distributions in $^3$He rather than proton
distributions in \trit{} and \het.

\begin{figure}[htpb]
\begin{center}
\includegraphics[width=3in]{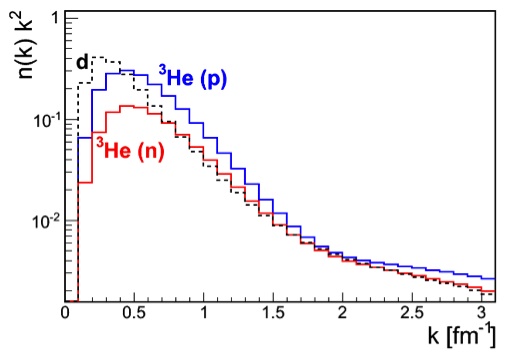}
\includegraphics[width=3in]{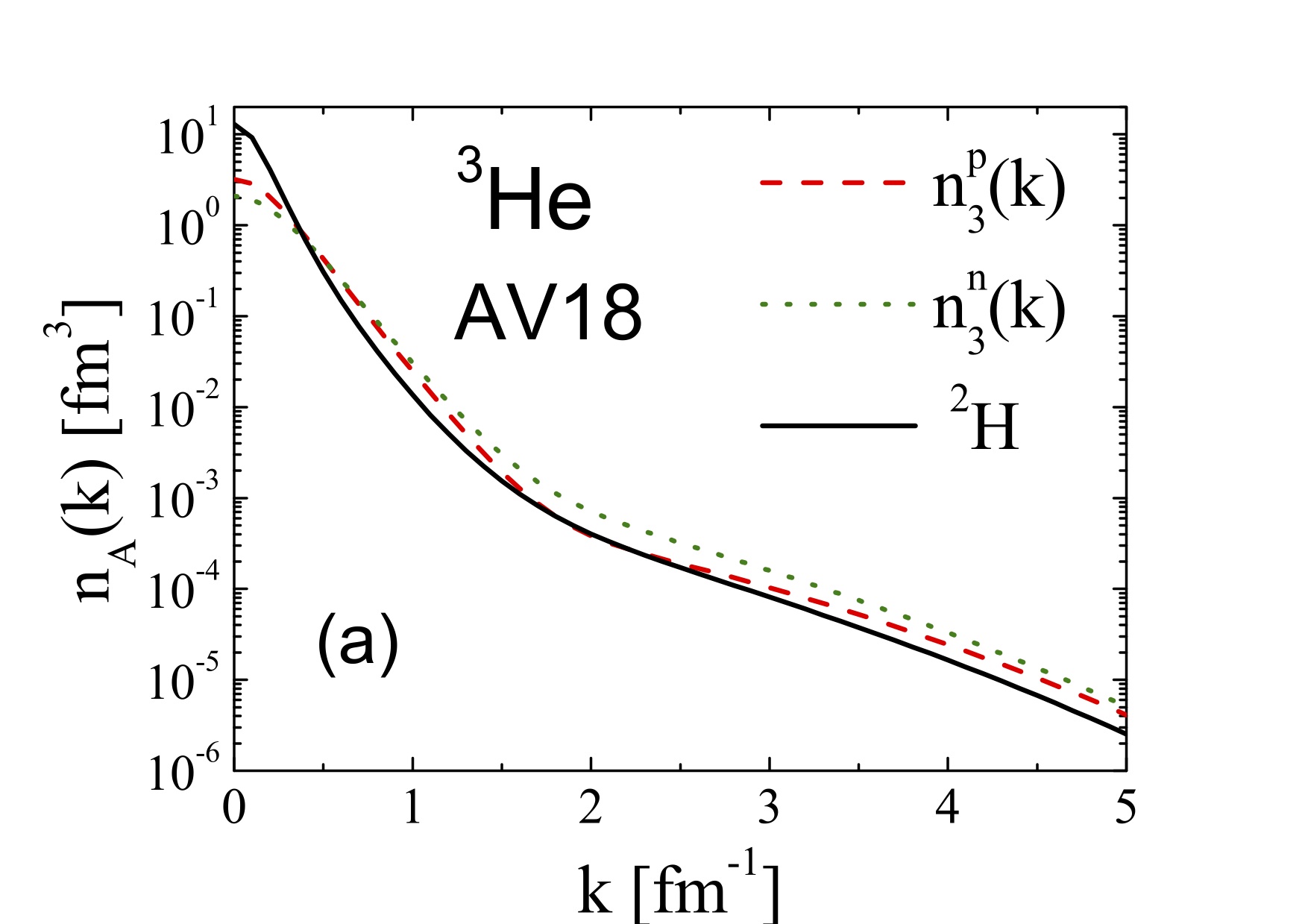}
\caption[]{\label{fig:MomDist} \baselineskip 13pt The momentum
  distributions of the proton and the neutron in \het{} and in
  deuterium.  Left: $k^2n(k)$ from \cite{wiringapc,wiringa95,veerasamy11,pieper01}.  The
  proton's momentum distribution integral is normalized to $Z=2$ and
  the neutron's to $N=1$. The deuterium distribution is multiplied by
  $a_2(\het/d)=2$, the ratio of the cross sections for \het$(e,e')$ to
  $d(e,e')$ at $1.5\le x\le 2$ \cite{fomin12}. Right: $n(k)$ from
  \cite{alvioli13}.  The integrals of the momentum distributions are
  normalized to one.}
\end{center}
\end{figure}

Fig.~\ref{fig:MomDist}   shows the nucleon momentum distribution in the
deuteron and  the proton
and neutron momentum distribution in \het{} calculated by
\cite{wiringapc,wiringa95,veerasamy11} using 
the Argonne V18 (AV18) $NN$ potential for the deuteron and
Variational Monte Carlo (VMC) calculations using the Argonne V18 + Urbana IX
(AV18+UIX) Hamiltonian \cite{wiringapc,pieper01} for \het.   The
proton momentum distribution in \het{} is significantly greater than
that of the neutron at low momenta but is approximately the same for
$1.5 \le k \le 2.5$ fm$^{-1}$ (corresponding to $300 \le p \le 500$
MeV/c).   Fig.~\ref{fig:MomDist} also
shows the same quantities calculated by Alvioli {\it et al.} \cite{alvioli13}.

%$2N$-SRC dominate the momentum distribution starting at about $p=275$
%MeV/c \cite{egiyan06}. 
%  The $np$-SRC dominance region is at about $300
% \le p \le 500$ MeV/c where the momentum distributions of the $n$ in
% \het, the $p$ in \het, and the deuteron are similar.  For $p\ge 500$
% MeV/c, the proton momentum distribution in \het{} increases, showing
% the effect of configurations where all three nucleons have high
% momentum (e.g., $3N$-SRC).

For momenta $300 \le p \le 500$ MeV/c, the nucleon momentum
distributions in \het, \trit, and the deuteron are all similar and
dominated by $np$ Short Range Correlations (SRC).  At higher momenta
(beyond the range of this experiment) the proton momentum distribution in \het{} increases
relative to that of the neutron due to 
three-nucleon-correlations (e.e., $ppn$ in \het).

\begin{figure}[htpb]
\begin{minipage}[b]{3. in}
\begin{center}
\includegraphics[width=3.3in]{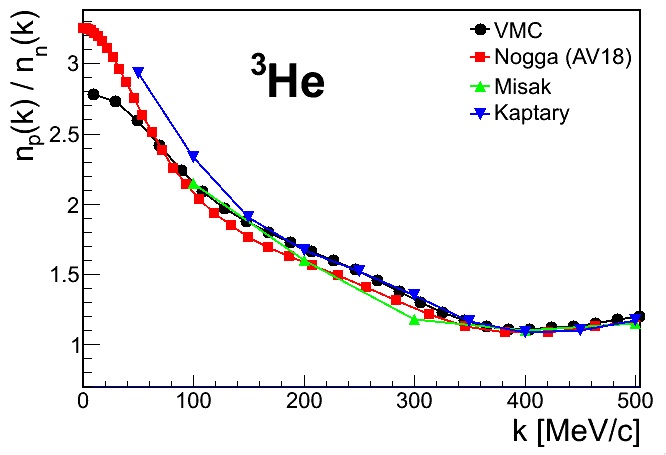}
\end{center}
     \end{minipage}\hfill
\vspace{-0.2in}
     \begin{minipage}[t]{3in}
\vspace{-2in}
\caption[]{\label{fig:MomDistRatio1} \baselineskip 13pt The calculated
  ratio of proton to neutron momentum distributions in \het. This
  corresponds to the ratio of momentum distributions for \het\eep{} to
  \trit\eep.}
\end{minipage}
%\end{center}
\end{figure}

The
naive nucleon-counting expectation is that the cross section ratio of
\het\eep{} to \trit\eep{} will equal two, the ratio of protons/neutrons in
\het.  However as shown in Fig. \ref{fig:MomDistRatio1}, the calculated ratio decreases rapidly from about three at low-momentum to about one at high
momentum.  Thus, by moving equal numbers of protons and neutrons from
the low momentum to the high momentum regions,  the correlations
increase the low momentum ratio from two to three and decrease the
high momentum ratio from two to one.

\vspace {1.2in}

\begin{figure}[htpb]
\begin{minipage}[b]{4. in}
\begin{center}
\includegraphics[width=4in]{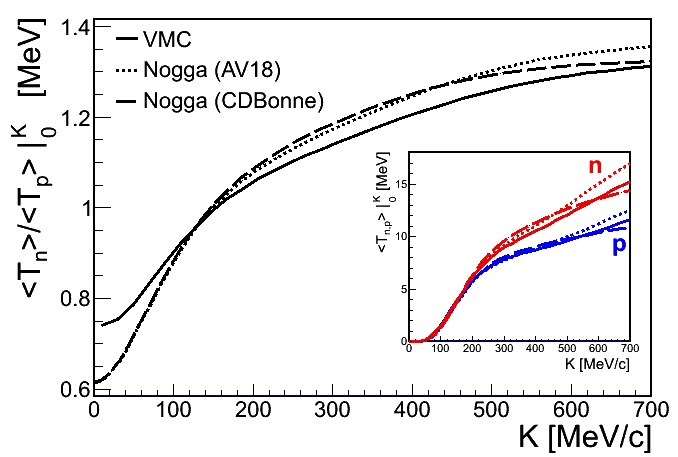}
\end{center}
     \end{minipage}\hfill
%\vspace{-.2in}
     \begin{minipage}[t]{2in}
\vspace{-2.5in}
\caption[]{\label{fig:TpRatio} \baselineskip 13pt The ratio of the
  neutron to proton average kinetic energy in \het{} integrated up to
  momentum $k$ as in Eq.{} \ref{eq:Tp}. Inset: The average kinetic
  energy per neutron (blue) and per proton (red) in \het{} integrated
  up to momentum $k$.  The neutron in \het{} corresponds to our
  proposed measurement of \trit\eep.}
\end{minipage}
%\end{center}
\end{figure}

The correlations also change the relative kinetic energies of protons
in \het{} and \trit{} (i.e., of the majority and minority nucleons in
the $A=3$ system).  We calculate the build-up of the average kinetic energy of the
proton (or neutron) up to some momentum $k$ as
\begin{equation}
\langle T_p\rangle\vert_0^k = \frac{\int_0^k
  n_p(k')(\sqrt{m_p^2+k'^2}-m_p)d^3k'}{\int_0^\infty n_p(k')d^3k'} \quad .
\label{eq:Tp}
\end{equation}
The majority ($p$ in \het) and minority ($p$ in \trit) average kinetic energies and their ratio are
shown in Fig. \ref{fig:TpRatio}.  Integrated up to momenta of about 150 MeV/c, the
majority have more kinetic energy; integrated up to higher momenta the ratio is
inverted and the minority have more kinetic energy.

\vspace {3.2in}

As we will discuss below, we can measure the \het\eep{} and \trit\eep{} cross
sections and their ratio very accurately.  We can use the ratio to very reliably extract the
ratio of ground state minority to majority nucleon momentum
distributions in $A=3$ nuclei  and the ratio of minority to majority average
kinetic energy with very small sensitivity to FSI.

\subsection{Relevance to other fields}

% {\bf how will our measurements affect any of these?}

\begin{figure}[htpb]
\begin{center}
\includegraphics[width=6in]{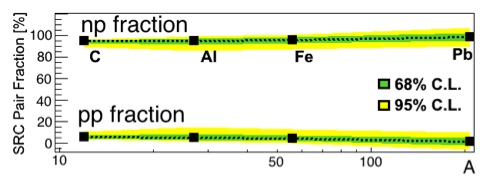}
\end{center}
\caption[]{\label{fig:NpFraction} \baselineskip 13pt The fraction of
  $np$ (top) and $pp$ (bottom) SRC pairs. The green and yellow bands
  reflect 68\% and 95\% confidence levels. $np$-SRC pairs dominate
  over $pp$-SRC in all measured nuclei.}
\end{figure}

In the symmetric $^4$He and $^{12}$C nuclei, every proton with
momentum $300 < k <600$
MeV/c (where $k_F \approx 220$ MeV/c) has a correlated partner nucleon
with neutron-proton ($np$) pairs outnumbering proton-proton ($pp$),
and by inference neutron-neutron ($nn$), pairs by a factor of
$\approx20$ \cite{subedi08,piasetzky06,korover14}. In a paper submitted
to Science we present data from Hall B analyzed as part of the data
mining project \cite{hen14}. We show, for the first time, the identification of SRC
pairs in the high-momentum tail of nuclei heavier than carbon and with
more neutrons than protons (i.e., $N>Z$). This publication
demonstrates clearly the universal nature of SRC pairs, which even in
heavy imbalanced nuclei such as lead ($N/Z = 126/82$) are still
dominated by $np$ pairs (see Fig.~\ref{fig:NpFraction}). This $np$-dominance causes
a greater fraction of protons than neutrons to have high momentum in neutron-rich
nuclei, thereby inverting the momentum sharing in imbalanced nuclei.

The $np$-dominance of SRC pairs and the resulting inversion of the
momentum sharing in heavy neutron rich imbalanced nuclei have wide
ranging implications in astro, nuclear and particle physics. These
include the determination of the density dependence of the nuclear
symmetry-energy up to supra-nuclear densities
\cite{Carbone12,Vidana11,Xu13,Xu11,Lattimer13}, analysis of
neutrino-nucleus scattering data for the determination of the nature
of the electro-weak interaction \cite{Fiorentini13a,Fiorentini13b},
and the isospin dependence of the EMC effect as a cause of the
standard-model NuTeV anomaly
\cite{Frankfurt88,Hen13,Weinstein11,Zeller02,Cloet09}.

One application is in neutrino physics where most experiments still
use a simple relativistic Fermi gas model to describe the
nucleus. Recent high precision measurements of charged current
quasi-elastic neutrino-nucleus scattering cross-sections
\cite{Fiorentini13a,Fiorentini13b} show the need to include the
effects of $np$-SRC pairs in both their reaction model and detector
response. This is expected to be a crucial ingredient in facilitating
the precision requirements of next generation neutrino experiments
\cite{Int13}.

The NuTeV anomaly is a three standard deviation difference from the
Standard Model prediction in the extraction of the electroweak-mixing
(Weinberg) angle from neutrino deep inelastic scattering from iron
\cite{Zeller02}. This anomaly might be due to an Isospin dependent EMC
effect in iron \cite{Cloet09}. The EMC effect implies modification of
bound nucleon structure \cite{kulagin10} and was recently shown to linearly
correlate with the number of SRC (high momentum) pairs in nuclei
\cite{Weinstein11}. If the EMC nucleon modification is dominated by
high momentum nucleons and protons have higher momentum than neutrons
in heavy asymmetric nuclei, then this would provide an alternative model
\cite{Cloet09} for an Isospin dependent EMC effect which can
quantitatively explain the NuTeV anomaly.

The nuclear symmetry energy describes how the energy per nucleon in
nuclear matter changes as a function of the proton fraction. While its
value at the nuclear saturation density is relatively well constrained \cite{Lattimer13}, its
density dependence is not, largely due to uncertainties in the tensor
component of the nucleon-nucleon interaction \cite{Xu13,Xu11}. Knowledge of
this density dependence at supra-nuclear densities is important for
different aspects of nuclear astrophysics and in particular neutron
stars \cite{Lattimer13}.
Recent calculations show that the inclusion of high-momentum tails,
dominated by tensor force induced $np$-SRC pairs, dramatically softens
the nuclear symmetry energy at supra-nuclear densities
\cite{Carbone12,Vidana11,Xu13,Xu11}. 

Generalizing our results to other systems, the behavior of two
component Fermi systems that interact with a large scattering length
is constrained by universal relations which involve ``the contact'', a
measure of the number of short distance unlike-fermion pairs
\cite{Tan08a,Tan08b,Tan08c}. Recent experiments with symmetric
two-spin-state ultra-cold atomic gases measured the
contact. % We suggest extending these
% measurements to asymmetric systems where the discussed inversion of
% the momentum sharing can be observed. The large flexibility of these
% systems will allow a comprehensive study of this inversion with the
% asymmetry, density, and the strength of the short range
% interaction. The results of such studies will increase our insight on
% the general behavior of interacting two-component asymmetric Fermi
% systems.

This measurement will not measure a specific parameter needed for
these different systems.  However, it will contribute to our general
understanding and will help constrain the theories common to this
measurement and to these systems.

\subsection{Impact on the 12 GeV JLab program}

The results of this proposed measurement will complement other
12 GeV JLab experiments, particularly EMC effect measurements of
light nuclei and inclusive \ee{} measurements of light nuclei.

E12-11-112 will measure inclusive electron scattering, \ee, from
\het{} and \trit{} from $x_B\approx 0.7$ to $x_B=3$, covering the
resonance, quasielastic, and $x_B>1$ regions.  Inclusive electron
scattering at $x_B>1$ is sensitive to the integral of the nuclear
momentum distribution from some threshold momentum to infinity, where
the threshold momentum increases with $x_B$.  Exclusive \eep{} cross
sections are sensitive to the momentum distribution $n(p)$ at
$p\approx p_{miss}$.  Thus, our proposed \eep{} exclusive electron
scattering measurement will complement the inclusive measurements.  In
addition, the theoretical uncertainties of the exclusive measurements
will be very different from those of inclusive measurements.  As this
will be our only opportunity to study \trit{} at Jefferson Lab, it is
important to perform both experiments.

The EMC effect is the deviation from
unity of the per-nucleon deep inelastic scattering (DIS) cross section of
nucleus $A$ relative to deuterium.  The EMC effect cannot be explained
by purely nucleonic effects (e.g., nucleon motion and binding energy)
and therefore implies that nucleons are modified in nuclei \cite{kulagin10}.

\begin{figure}[htpb]
\begin{center}
\includegraphics[width=2.5in]{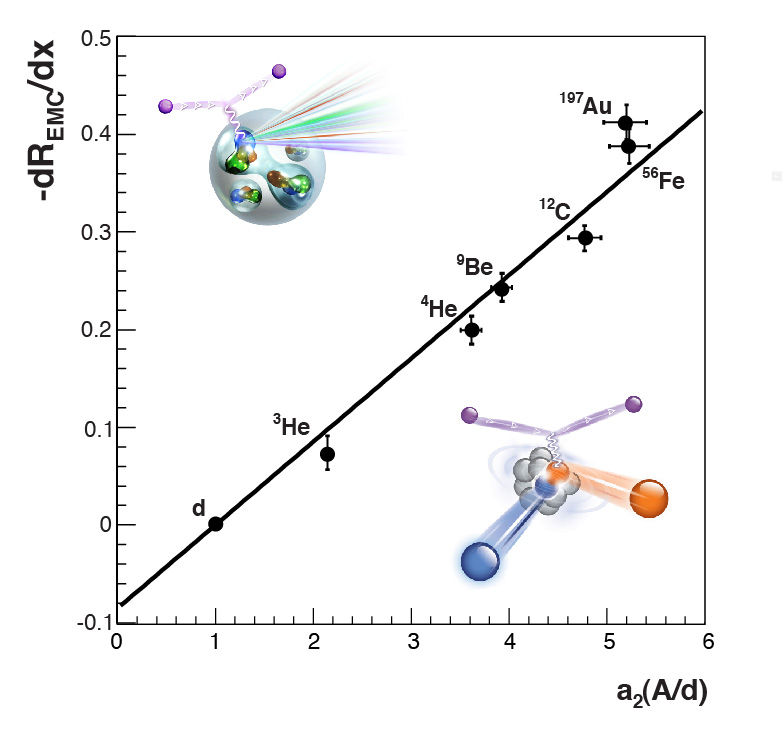}
\caption[]{\label{fig:EmcSrcEbind} \baselineskip 13pt Linear correlation between
  the strength of the EMC effect and the amount of 2N-SRC in nuclei
  \cite{hen12}. 
}
\end{center}
\end{figure}

The magnitude of the EMC effect is directly proportional to the number of
SRC pairs \cite{Weinstein11,hen12}  (see Fig.~\ref{fig:EmcSrcEbind}) and thus
the EMC effect is associated with large momentum (large
virtuality) nucleons in nuclei.  

The MARATHON experiment will measure the ratio of DIS cross sections
in \het{} and \trit{} to determine the ratio of $F_2^p/F_2^n$.  They
expect that off-shell effects due to the modification of the bound
nucleon structure will be small.
However, since models predict that the modification of the bound nucleon structure
function is proportional to nucleon virtuality
\cite{kulagin10} and since protons in \trit{} are expected to have higher momentum
than protons in \het{}, this means that $F_2^p$ will not be the same
in \het{} and \trit{} (and similarly for neutrons and $F_2^n$).

By directly measuring the proton  (and by inference the neutron)
momentum distributions in \het{} and \trit, this experiment will
provide information to help correct the MARATHON data for possible
off-shell effects.

\subsection{Previous Measurements}

The most direct way of studying nucleon momentum distributions in
nuclei is to measure
the quasi-elastic (QE) $^3$He\eep{} and $^3$H\eep{} reactions as a function
of missing momentum, $\vec p_m = \vec q - \vec
p_p$, where $\vec p_p$ is the momentum of the outgoing, observed
proton and $\vec q$ is the momentum transfer. The missing momentum
$\vec p_m$ equals $\vec p_r$,  the momentum of the $A-1$ recoil.  Within the Plane Wave
Impulse Approximation (PWIA) where Final State Interactions (FSI) are
neglected, $\vec p_{init} = -\vec p_m$ where $\vec p_{init}$ is the initial momentum of
the target nucleon before the interaction.

\begin{figure}[htpb]
\begin{center}
\includegraphics[width=3in]{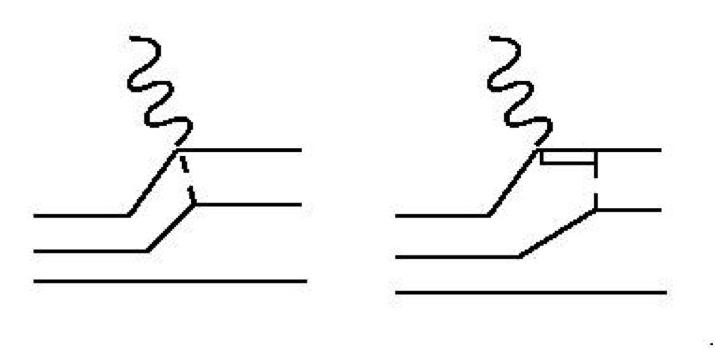}
\caption[]{\label{fig:TwoBodyCurrents} \baselineskip 13pt Diagrams for
Meson Exchange Currents (left) and Isobar Configurations (right).}
\end{center}
\end{figure}

However, depending on the selected kinematics, other reaction
mechanisms can contribute to the cross section. The outgoing (struck)
proton can rescatter from the other nucleons (FSI), or the virtual
photon can couple to the exchanged meson (MEC)  or the virtual
photon can excite the nucleon to an intermediate $\Delta$ isobar state (IC).
Fig.{} \ref{fig:TwoBodyCurrents} shows schematically how IC and MEC
can cause the measured missing momentum to be different from the
genuine ground state  momentum distribution.

\begin{figure}[htpb]
\begin{center}
\includegraphics[width=3in]{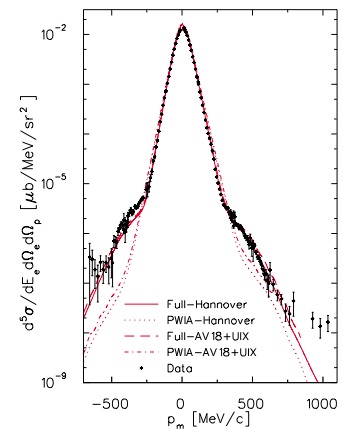}
\includegraphics[width=3in]{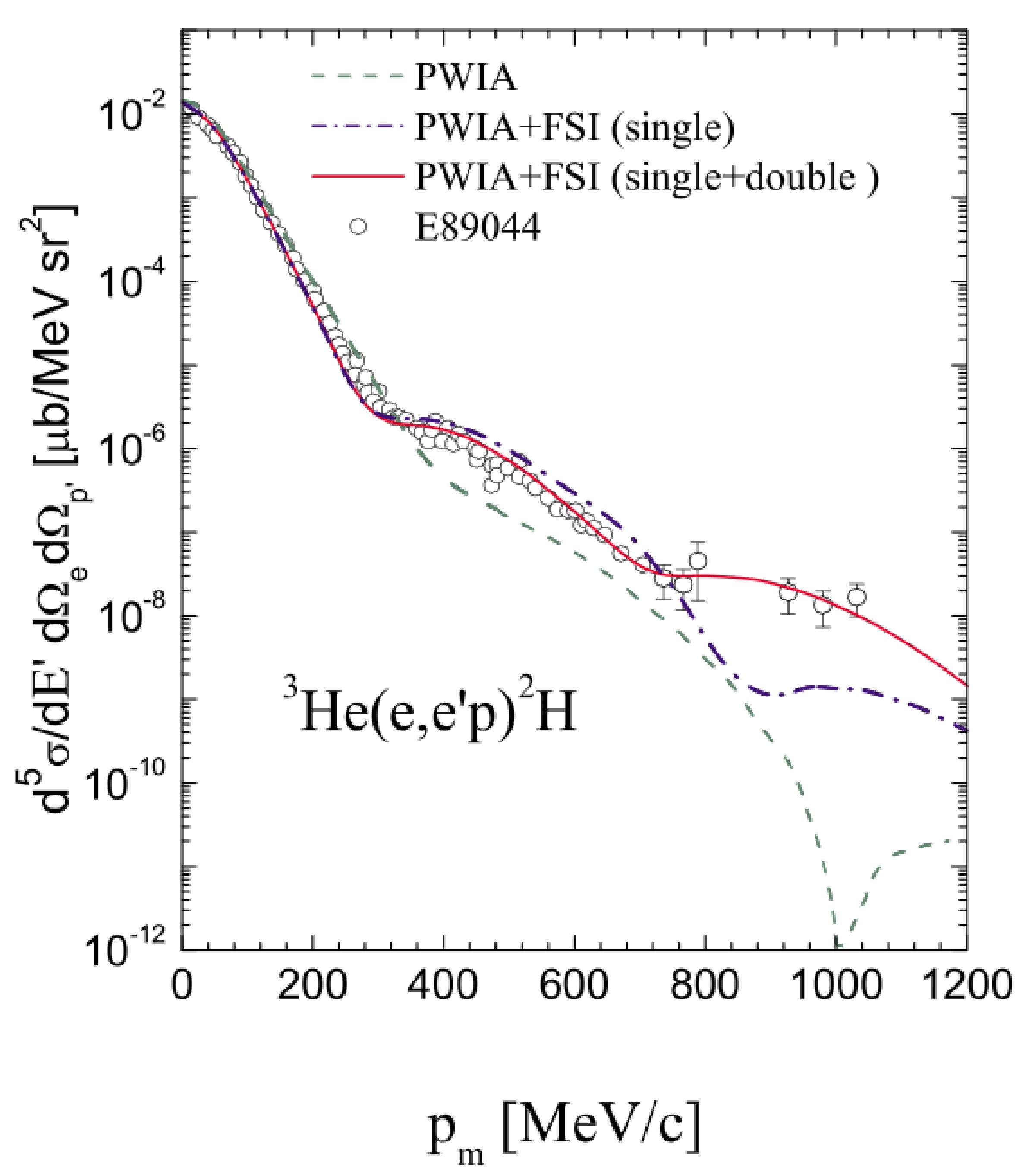}
\caption[]{\label{fig:HeEep2bbu} \baselineskip 13pt The measured
  \het\eep$^2$H cross section as a function of the missing momentum
  $p_m$.  The left figure also shows PWIA and full calculations in the
  diagrammatic approach by Laget for two different ground state wave
  functions (see \cite{rvachev05} and references therein). The right
  figure shows the same data with calculations by Ciofi degli Atti and
Kaptari \cite{cda05}. The dashed line corresponds to the PWIA, the
dot-dashed line includes FSI with single rescattering and the solid
line includes both single and double rescattering \cite{cda05}.}
\end{center}
\end{figure}

\begin{figure}[htpb]
\begin{center}
\includegraphics[width=3in]{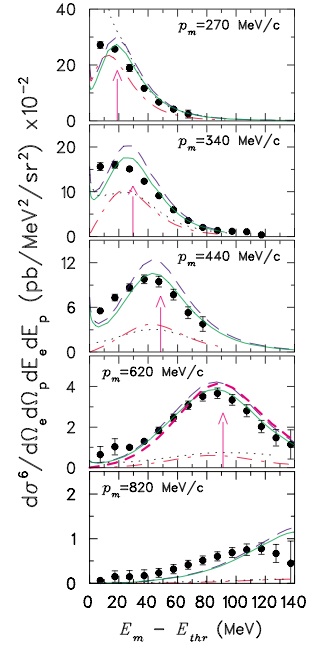}
\includegraphics[width=3in]{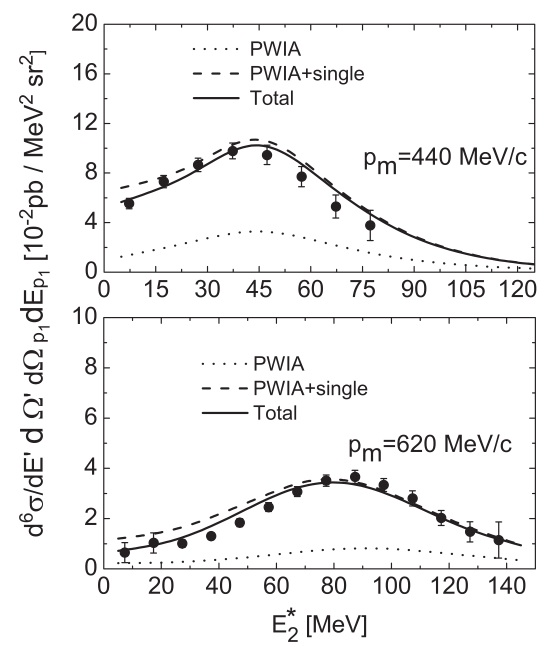}
\caption[]{\label{fig:HeEepCont} \baselineskip 13pt (left) The cross section
  for the \het\eep$pn$ reaction as a function of missing energy \cite{benmokhtar05}.  The
  vertical arrow gives the peak position expected for disintegration
  of correlated pairs, $E_m = \omega - T_p - T_r = p_m^2/4m$ (where
  $T_p$ is the kinetic energy of the detected proton and $T_r =
  p_m^2/2m_{A-1}$ is the kinetic energy of the recoiling $A-1$
  ``nucleus''). The black dotted curve presents a PWIA calculation
  using Salme's spectral function and $\sigma_{cc1}$ electron-proton
  off-shell cross section and the red dash-dotted line is a Laget PWIA
  calculation. Other curves are Laget's calculations for PWIA+FSI
  (black long dashed line) and his full calculation (solid green
  line), including meson-exchange currents and final-state
  interactions. In the 620 MeV/c panel, the additional bold red short
  dashed curve is a calculation with PWIA + FSI only within the
  correlated pair \cite{laget03,laget05}. (right) The same
  experimental cross section plotted vs $E_2^* = E_m - E_{thr}$
  compared to the unfactorized calculations of Alvioli,
  Ciofi degli Atti and Kaptari \cite{alvioli10}.  The dotted line
  shows the PWIA calculation, the dashed line includes FSI with single
  rescattering and the solid line includes FSI with both single and
  double rescattering.}
\end{center}
\end{figure}

Due to these competing reaction channels, previous experiments at $Q^2
< 1$ (GeV/c)$^2$
\cite{bussiere81,blomqvist98,boeglin08,ulmer02,rvachev05} did not
strongly constrain the high momentum components of the ground state
momentum distribution.  The Jefferson Lab \het\eep{} measurement was
performed in kinematics such that the cross sections measured at high-missing-momentum
($p_m \ge 300$ MeV/c) were dominated by FSI (see Figs.{}
\ref{fig:HeEep2bbu} and \ref{fig:HeEepCont})
\cite{rvachev05,benmokhtar05}.  The FSI domination is shown by the
significant disagreement between PWIA calculations and the measured
cross sections and by the relative agreement between calculations
including FSI and those same cross sections.  This FSI domination is
not surprising, because small angle rescattering of the knocked-out
proton contributes significantly at some kinematics.

\begin{figure}[htpb]
\begin{center}
\includegraphics[width=3in]{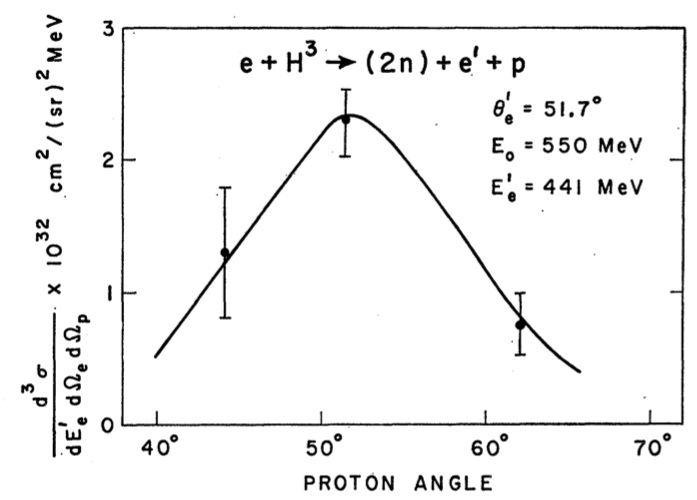}
\caption[]{\label{fig:Heep1964} \baselineskip 13pt  The cross section
  for \trit\eep{} as a function of proton angle \cite{johansson64}.}
\end{center}
\end{figure}

The only previous measurement on \trit\eep{} dates from 1964
\cite{johansson64}.  Unfortunately, little can be learned about the
proton momentum distributions in \trit{} from this measurement, due to
its low momentum transfer and limited statistics (see Fig.~\ref{fig:Heep1964}).

\subsection{Minimizing Final State Interactions \label{sec:FSI}}

\begin{figure}[htpb]
\begin{center}
\includegraphics[width=5in]{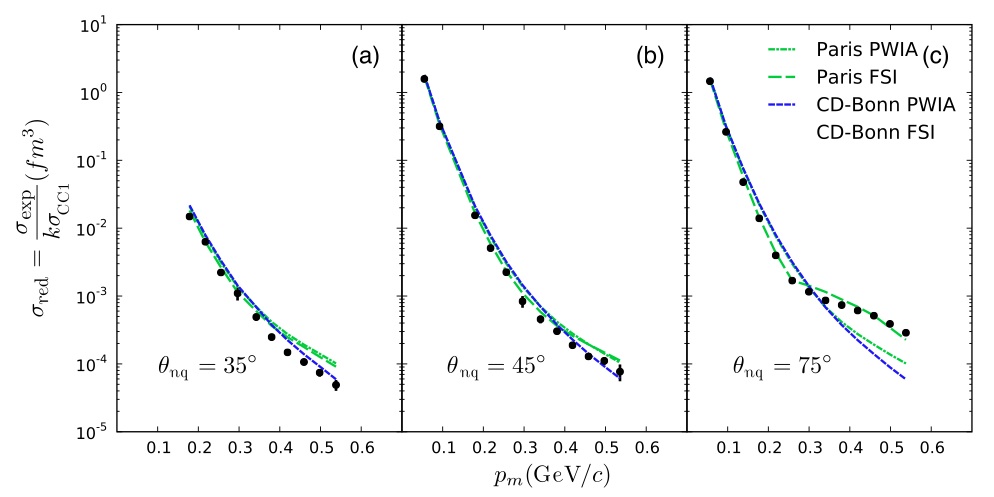}
\caption[]{\label{fig:ThetaRQDeep} \baselineskip 13pt The reduced
  cross section for \deep{} as a function of missing momentum for
  three different lab recoil angles, (a) $\theta_{rq}=35^\circ$, (b)
  $\theta_{rq}=45^\circ$, and (c) $\theta_{rq}=75^\circ$.  All
  calculations are by M. Sargsian \cite{boeglin11}. }
\end{center}
\end{figure}

Fortunately, measurements on the deuteron show that we can select
kinematics to minimize the effects of FSI \cite{boeglin11}.  Fig.{}
\ref{fig:ThetaRQDeep} shows that the impact of FSI on the cross section
decreases rapidly as $\theta_{rq}$, the angle between the recoil momentum
($\vec p_{recoil} = \vec p_m$)   and $\vec q$ in the laboratory frame, decreases from 75$^\circ$ to
35$^\circ$.

\begin{figure}[htpb]
\begin{minipage}[b]{3.5 in}
\begin{center}
\includegraphics[width=3.5in]{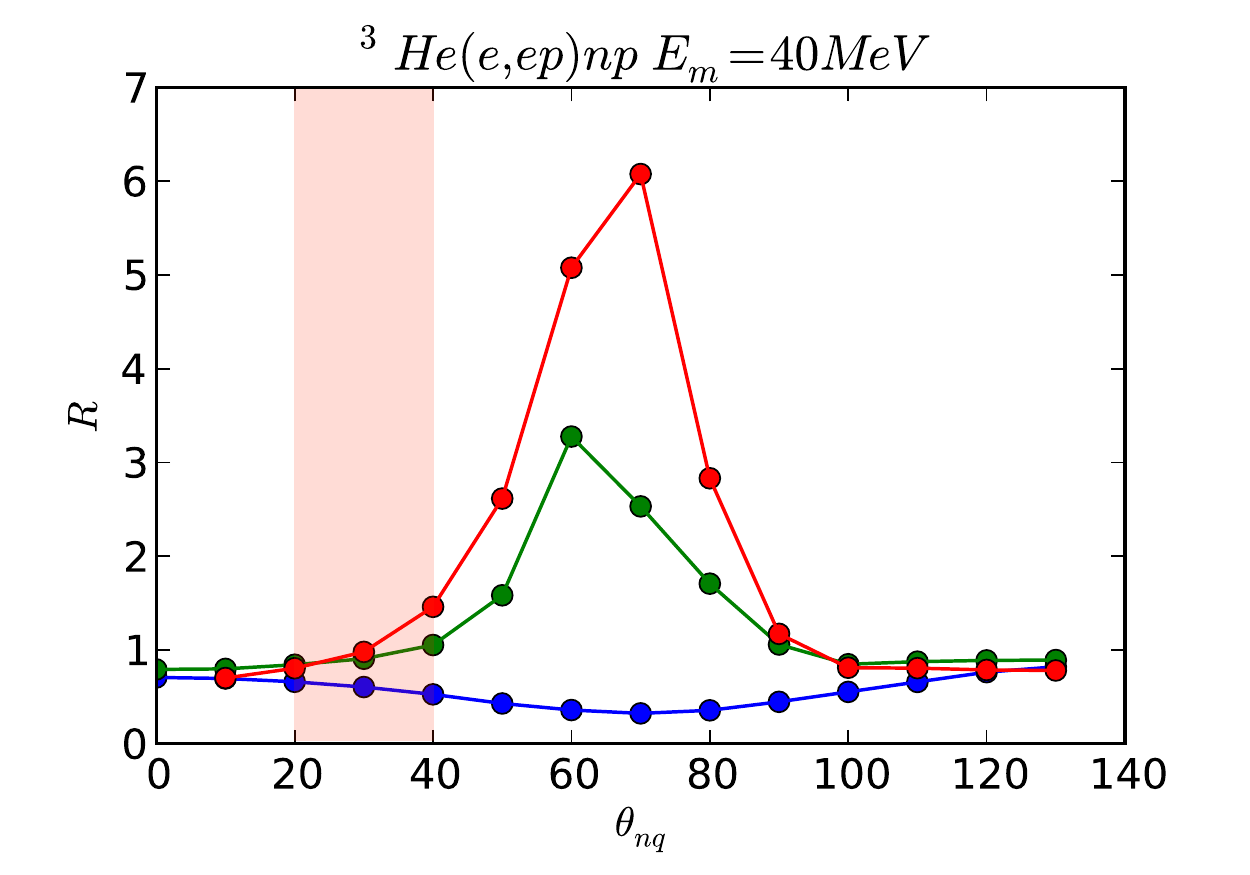}
\end{center}
     \end{minipage}\hfill
%\vspace{-1.2in}
     \begin{minipage}[t]{2.5in}
\vspace{-2in}
\caption[]{\label{fig:3HeFsi} \baselineskip 13pt The calculated
  \het\eep{}  ratio of the cross section
which includes rescattering of the struck nucleon (FSI) to the
PWIA cross section for $p_m = 0.2$ (blue), 0.4 (green), and 0.5 (red) GeV/c as a
function of $\theta_{rq}$, the angle between the recoil momentum and
$\vec q$  in the laboratory frame \cite{misakpc}. The tan band
indicates the angles for the measurements proposed here.}
\end{minipage}
%\end{center}
\end{figure}

We expect the same FSI suppression to hold for the nucleons in the
correlated pair in \het{} and \trit. Calculations by Ciofi degli Atti
\cite{cdapc} and by Sargsian \cite{misakpc} show that FSI are
minimized at smaller values of $\theta_{rq}$.  Fig.~\ref{fig:3HeFsi}
shows that the effects of FSI peak at $\theta_{rq}\approx 70^\circ$
and are much smaller for $\theta_{rq} \le 40^\circ$.

%\vspace{1.2in}
 
\begin{figure}[htpb]
\begin{center}
\includegraphics[width=3.2in]{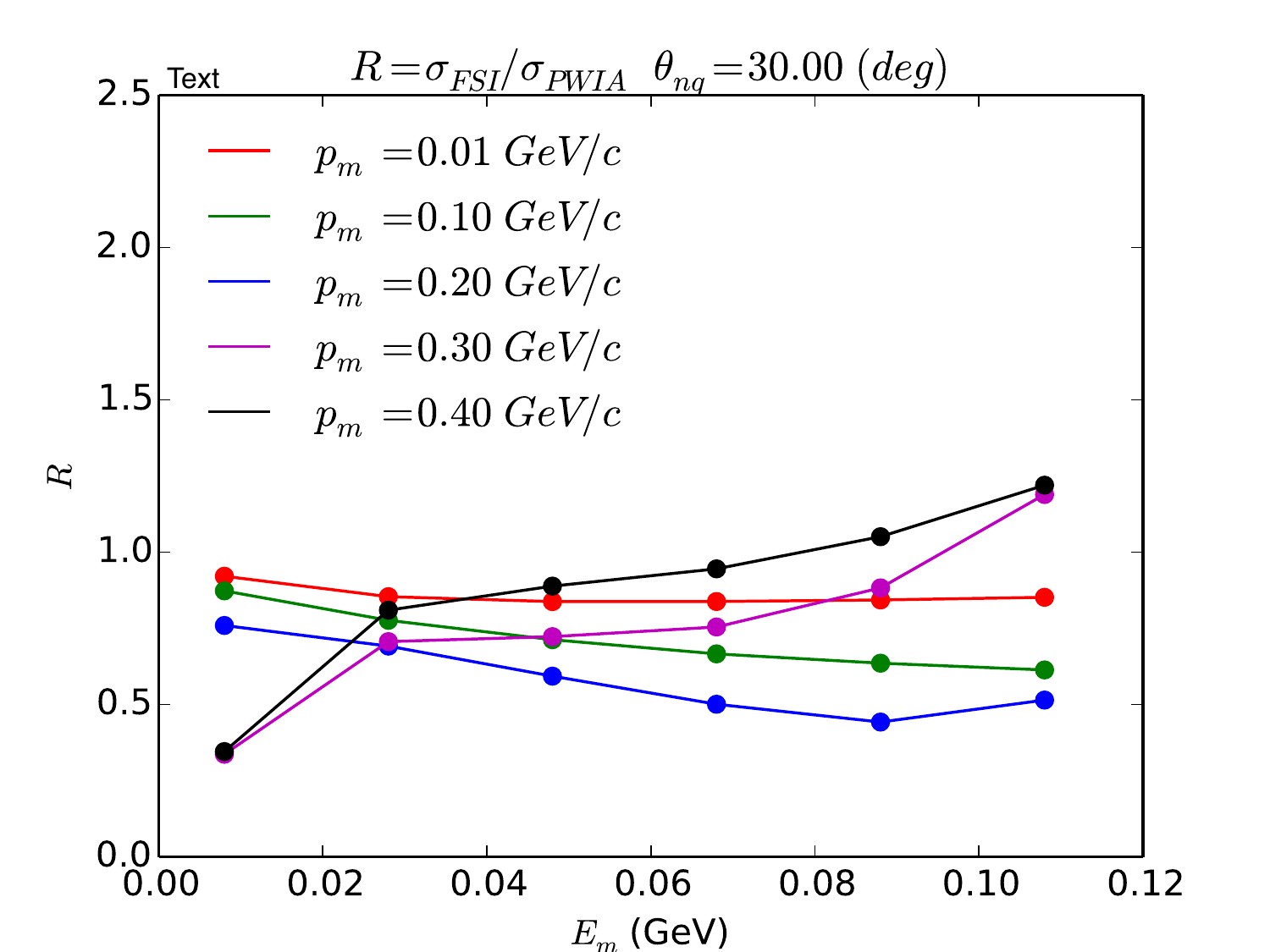}
\includegraphics[width=3.2in]{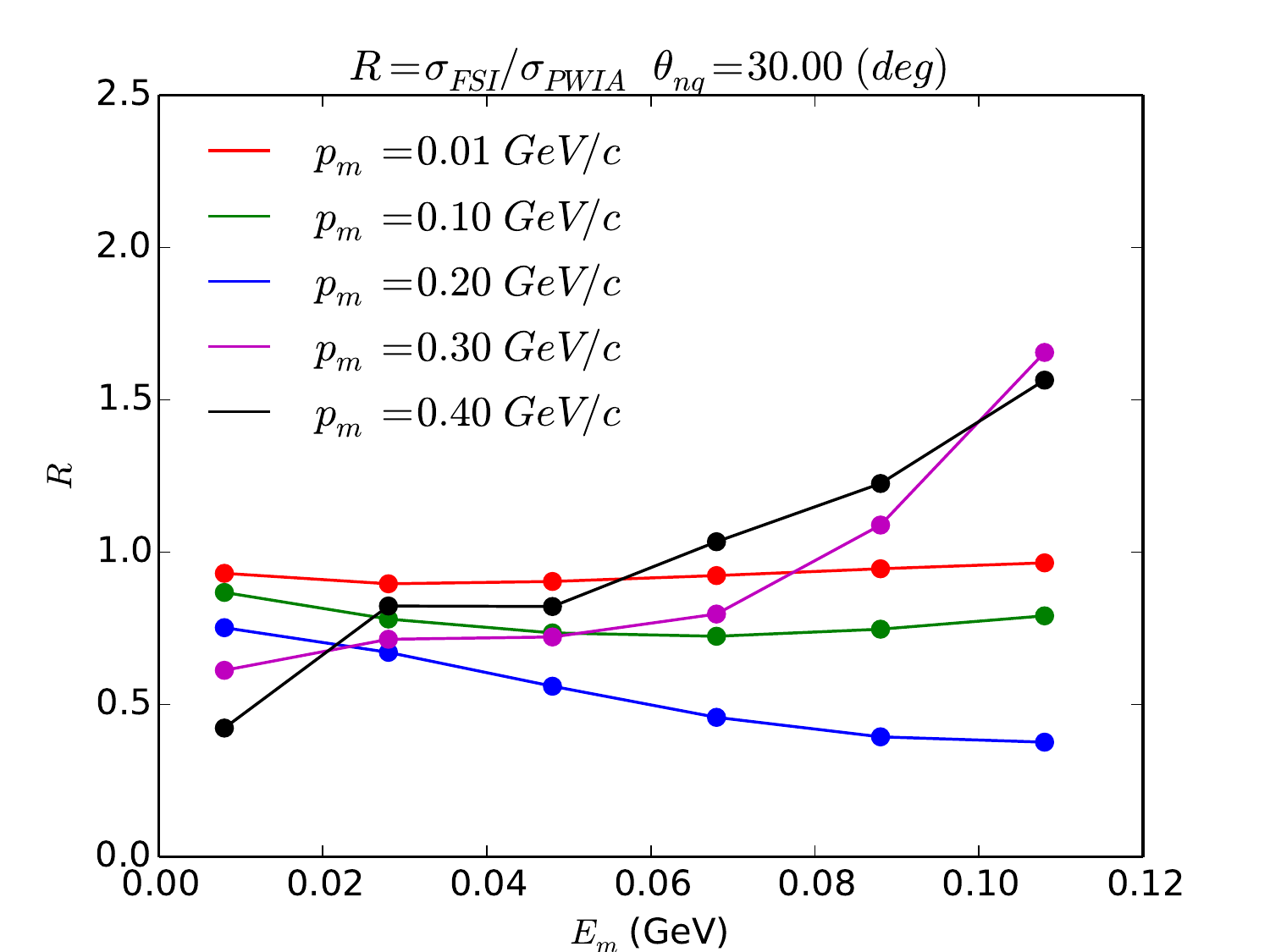}
\caption[]{\label{fig:3HFsiPwiaRatio30} \baselineskip 13pt The ratio of the FSI
calculation which includes rescattering of the struck nucleon to the
PWIA cross section as a function of missing
energy $E_m = \omega - T_p - T_r$ for 
$\theta_{rq} = 30^\circ$ and for various missing momenta as calculated
by Sargsian
\cite{misakpc}. Left: \trit; Right: \het.}
\end{center}
\end{figure}

Fig.~\ref{fig:3HFsiPwiaRatio30} shows the calculated ratio of the FSI 
to  PWIA \trit{} and \het\eep{} cross sections for a range of missing energies and missing
momenta at a recoil angle $\theta_{rq}=30^\circ$.  The ratios are
almost all reasonably close to unity.

% \vspace {1.2in}

% \begin{figure}[htpb]
% \begin{minipage}[b]{3.5 in}
% \begin{center}
% \includegraphics[width=3.5in]{./3H_FSI_PWIA_Em_30.pdf}
% \end{center}
%      \end{minipage}\hfill
% %\vspace{-1.2in}
%      \begin{minipage}[t]{2.5in}
% \vspace{-2in}
% \caption[]{\label{fig:3HFsiPwiaRatio30} \baselineskip 13pt The ratio of the FSI
% calculation which includes rescattering of the struck nucleon to the
% PWIA cross section as a function of missing
% energy $E_m = \omega - T_p - T_r$ for 
% $\theta_{rq} = 30^\circ$ and for various missing momenta \cite{misakpc}. }
% \end{minipage}
% %\end{center}
% \end{figure}
% % \vspace {1.2in}

We will significantly reduce the remaining effects of FSI by forming
the ratio of \het\eep{} to \trit\eep.  In the simplest picture, the
struck proton has the same probability to rescatter from the other two
nucleons whether it was knocked out of \het{} or \trit.  

% In practice,
% calculations by Kaptari \cite{kaptaripc} show that in perpendicular
% kinematics at $p_m = 500$ MeV/c, where FSI increase the PWIA cross
% section by a factor of about four, they change the ratio of \het\eep{}
% to \trit\eep{} by only 20--30\% at the peak of the cross section (see
% Fig.~\ref{fig:PerpKinPWIAFSI}).  However, in perpendicular kinematics,
% the effects of FSI are additive because the rescattering shifts
% strength from smaller to larger missing momentum.   Therefore, even
% though the PWIA and PWIA+FSI cross section ratios are similar, it is
% important to measure the cross section away from perpendicular
% kinematics to reduce the effect of FSI on both the cross sections and
% their ratios.

% \begin{figure}[htpb]
% \begin{center}
% \includegraphics[width=3in]{./X-sect.jpg}
% \includegraphics[width=3in]{./Ratio.jpg}
% \caption[]{\label{fig:PerpKinPWIAFSI} \baselineskip 13pt (left) The
%   \het\eep{} and \trit\eep{} cross sections in perpendicular
%   kinematics (kinematics very similar to that of the data in Fig.{}
%   \ref{fig:HeEepCont}) calculated by Kaptari for
%   both PWIA and for PWIA plus FSI.  (right) The ratio of \het\eep{} to
%   \trit\eep{} for the PWIA and PWIA+FSI calculations.  Note that, even
%   at this kinematic point where FSI change the cross section by a
%   factor of almost four, the effect of the FSI on the \het/\trit{}
%   ratio is much smaller \cite{kaptaripc}. }
% \end{center}
% \end{figure}

\begin{figure}[htpb]
\begin{center}
\includegraphics[width=5in]{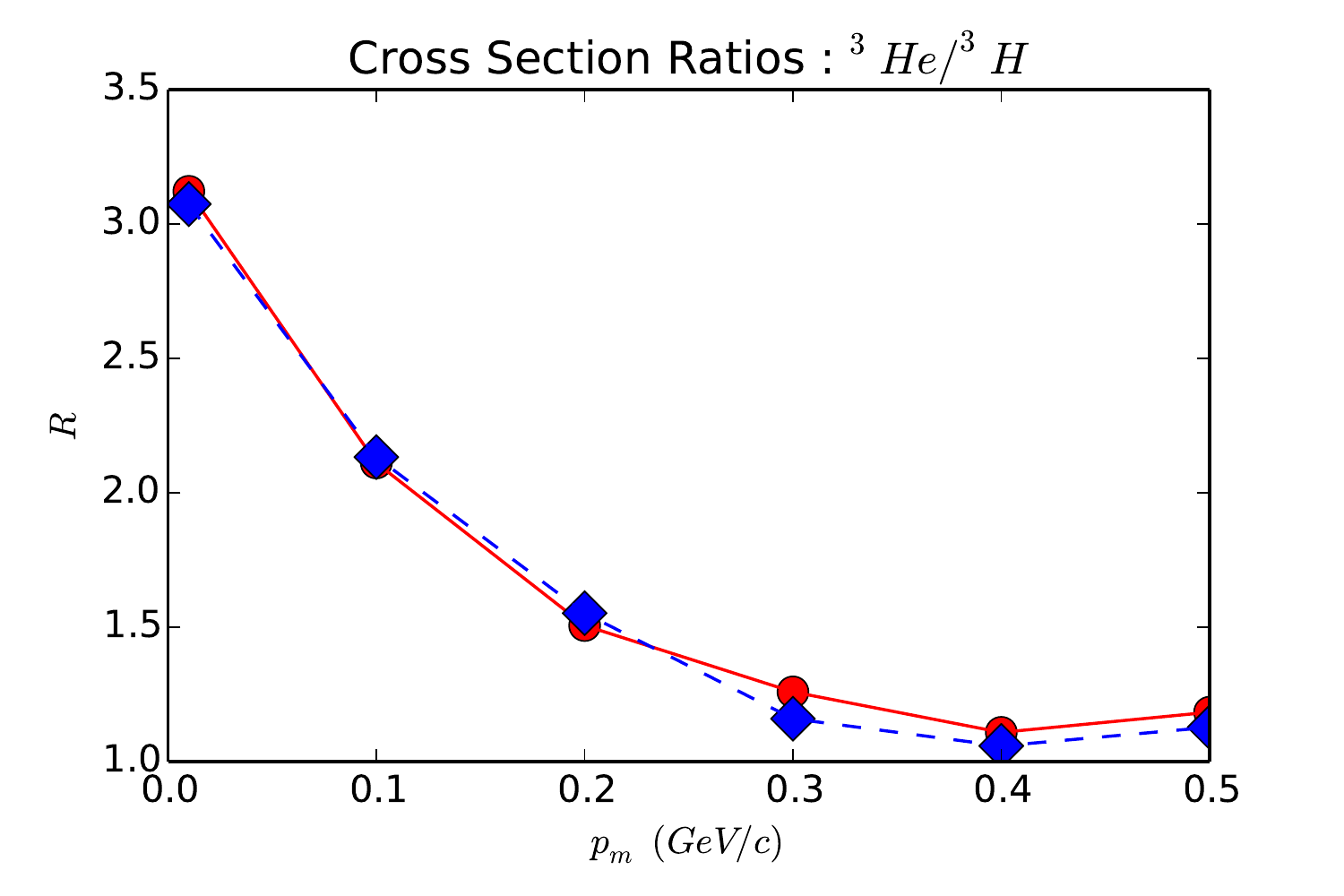}
\caption[]{\label{fig:3He3HFsiPwiaRatio} \baselineskip 13pt The ratio
  of the \het\eep{} to \trit\eep{} cross sections integrated over
  missing energy at $Q^2=2$ (GeV/c)$^2$
  and $\theta_{rq}=30^\circ$ for  PWIA calculations (red solid curve)
  and FSI calculations (blue  dashed curve) by Sargsian \cite{misakpc}. }
\end{center}
\end{figure}

Calculations by Sargsian at our proposed kinematic settings of $Q^2=2$ (GeV/c)$^2$ and
$\theta_{rq}=30^\circ$ show that the effects of FSI almost entirely cancel in the
ratio of \het\eep{} to \trit\eep{} cross sections (see Fig. \ref{fig:3He3HFsiPwiaRatio}).  

% The effects of FSI only partially cancel in the ratios of
% \het\eep/\deep{} and \trit\eep/\deep{} (see
% Fig.~\ref{fig:3He3HDFsiPwiaRatio}).  We will use the measured
% \het\eep/\deep{} and \trit\eep/\deep{} cross section ratios to test our theoretical
% understanding of final state interactions in these simplest of nuclei.

% \begin{figure}[htpb]
% \begin{center}
% \includegraphics[width=5in]{./3He_3H_D_FSI_PWIA_30.pdf}
% \caption[]{\label{fig:3He3HDFsiPwiaRatio} \baselineskip 13pt The ratio
%   of the \het\eep{} (blue) and \trit\eep (green) to \deep{} cross sections integrated over
%   missing energy at $Q^2=2$ (GeV/c)$^2$
%   and $\theta_{rq}=30^\circ$ for  PWIA calculation (dashed curve)
%   and FSI calculation (solid curve) \cite{misakpc}. }
% \end{center}
% \end{figure}

\section{The measurement}

We propose to measure the \het\eep{} and \trit\eep{} cross sections
for $0\le p_m \le 500$ MeV/c using the MARATHON target and the Hall A
HRS spectrometers in their standard configuration in order to extract information about the
ground state momentum distributions. 

\begin{figure}[htpb]
\begin{center}
\includegraphics[width=3in]{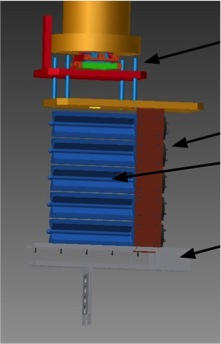}
\includegraphics[width=3in]{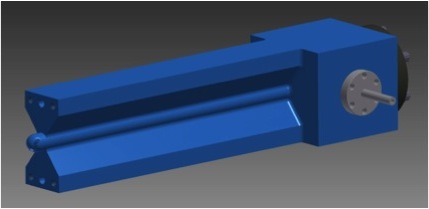}
\caption[]{\label{fig:target} \baselineskip 13pt The MARATHON target.
 (left) The four identical gas target cells are shown in blue and the optics
  target and solid targets are shown in grey. The beam is incident
  from the right.  (right) A detailed view
  of one target cell. 
}
\end{center}
\end{figure}

The MARATHON target has four identical 25-cm sealed-cell gas target
cells, with a maximum room-temperature pressure of 200 psi for T$_2$
and 400 psi for H$_2$, D$_2$ and $^3$He (see Fig.~\ref{fig:target}).
Each target cell is sealed and is cooled during target operation.  The open cell design allows a wide range of
  scattering angles.  The wall thickness is 0.018'' Al (120 mg/cm$^2$)
  and the entrance and exit windows are 0.010'' Al (65 mg/cm$^2$). The
target can withstand a maximum beam current of 25 $\mu$A.  The
monatomic \het{} and diatomic \trit{} (T$_2$) targets will have the
same thickness (82 mg/cm$^2$) and number density.  This gives a
maximum luminosity of $7.4\times10^{36}$ nucleons/(cm$^2\cdot$s), much
less than the maximum Hall A luminosity of $10^{39}$
nucleons/(cm$^2\cdot$s).  At this low luminosity, backgrounds and
other rate-related effects should be negligible.  The proton
spectrometers will not see the entrance and exit windows.

 In order to minimize the effects
of competing reaction channels (MEC, IC, and FSI) and to maximize our
sensitivity to the ground state momentum distribution, we will measure
at
\begin{itemize}
\item high $Q^2$ ($Q^2 \approx 2$ (GeV/c)$^2$)
\item $x=Q^2/2m\omega > 1$
\item small $\theta_{rq}$ ($\theta_{rq} < 40^\circ$)
\end{itemize}
In addition, we will measure the different nuclei at identical
kinematic settings with the identical experimental setup.
This will make it possible to probe the genuine momentum distributions
with significantly smaller experimental and theoretical corrections
and hence uncertainties.

Since, at fixed $Q^2$, cross sections increase with beam energy, we will measure the
cross section with two-pass beam, $E_0 = 4.4$ GeV.  This is the
highest beam energy we can use and still detect the scattered electron
in the Hall A HRS.

As $Q^2$ increases, the cross section decreases and the effects of IC
and MEC decrease.  In addition, the accuracy of the Eikonal approximation
for calculating the effects of FSI increases.  We will measure at
$Q^2=2$ (GeV/c)$^2$ as a compromise between decreasing cross
section and increasing ease of interpretation.

Based on deuteron results \cite{boeglin11} and \het{} calculations
\cite{cdapc,misakpc}, we plan to reduce the effects of FSI on the
cross sections by measuring the \eep{} reaction at an angle of
approximately $\theta_{rq} \approx 30^\circ$ between the nuclear
recoil and the momentum transfer.  

Due to the small $Z$ of the target nuclei, radiative effects will be
significantly smaller than in previous Hall A \eep{} measurements of
$^{16}$O and $^{208}$Pb.  They will further cancel in the ratio of the \het{}
and \trit{} cross sections.  Coulomb corrections will be similarly
small.  Radiative effects will be calculated using the standard Hall A
methods of \eep{} experiments.

We selected the
central angle and momentum of the detected electron and knocked-out
proton  according to 2-body kinematics for proton knockout from a  deuteron.  For $A=3$ targets, this centers
the kinematics close to  the SRC missing energy peak where the 
cross section is maximum (see Fig.{} \ref{fig:HeEepCont}).  It also
ensures that we will measure both the two-body-break-up and
three-body-break-up channels of \het\eep.

The minimum in
the \het\eep{} to \trit\eep{} cross section ratio (i.e., where
$NN$-SRC dominate) should be at about 0.4 GeV/c (see Fig.{}
\ref{fig:MomDistRatio1}).
We will measure the \het/\trit{} ratio out to
$p_m\approx 0.5$ GeV/c.   Table
\ref{tab:kin} shows the kinematic settings to cover this range of
missing momentum.

We request one day of commissioning and calibration time to measure
spectrometer optics, pointing, $^1$H\eep, and one low-$p_{miss}$
\het\eep{} data point to overlap with
Refs.~\cite{rvachev05,benmokhtar05}.  We expect that the systematic
uncertainties due to errors in beam energy, beam charge measurements,
detector efficiencies and target thickness will be very similar to the
4.5\% reported in \cite{boeglin11} for \deep.

\begin{table}[hb]
\begin{center}
{\baselineskip 13pt
\begin{tabular}{|c|c|c|c|c|c|c|} \hline
$<p_m>$ & $x$ & $E_e$ & $\theta_e$ & $p_p$ & $\theta_p$ &
total beam \\
(MeV/c) && (GeV) &&(GeV/c)&&  time (days) \\ \hline
&\multispan4 Calibration and commissioning \hfil && 1  \\ \hline
100 & 1.15 & 3.47 & 20.86$^\circ$ & 1.607 & 48.67$^\circ$ & 1 \\ \hline
300 & 1.41 & 3.64 & 20.35$^\circ$ & 1.352 & 58.55$^\circ$ & 8 \\ \hline
&\multispan4 Total beam time request \hfil && 10  \\ \hline
\end{tabular}
}
\end{center}
\vspace{-13 pt}
\caption{\baselineskip 13pt  The central kinematics and beam time for each setting.
  The beam energy is 4.4 GeV and $Q^2=2.0$ (GeV/c)$^2$ for all
  settings.  
}
\label{tab:kin}
\end{table}

\subsection{MCEEP Simulations}
We calculated cross sections and rates using the deuteron PWIA
cross section integrated over the experimental acceptances using
MCEEP, including radiative corrections.  We used
the parameters of the MARATHON target, with a 25-cm long target cell,
a deuterium target density of 75 mg/cm$^2$, and a maximum beam current
of 25 $\mu$A.  The \het{} and \trit{} targets will have approximately
the same mass density as the deuterium target and will therefore have
about 2/3 of the number density.  We expect the \het\eep{} and \trit\eep{}
cross sections to be at least twice as large as the \deep{} cross section
at large missing momenta, compensating for the decreased number
density of the \het{} and \trit{} targets and for other expected
experimental inefficiencies.

\begin{figure}[htpb]
\begin{center}
\includegraphics[width=3in]{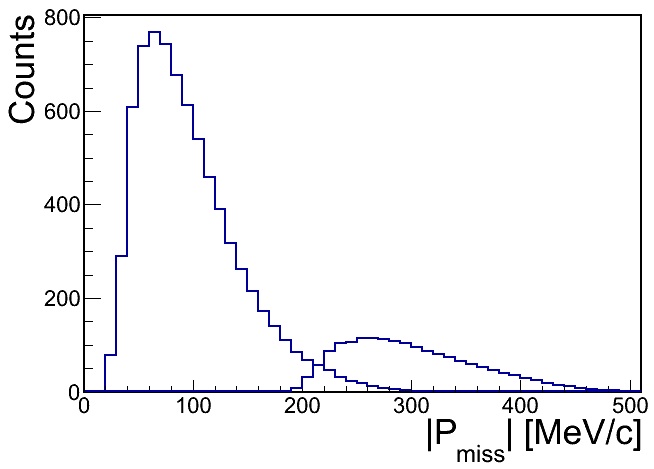}
\includegraphics[width=3in]{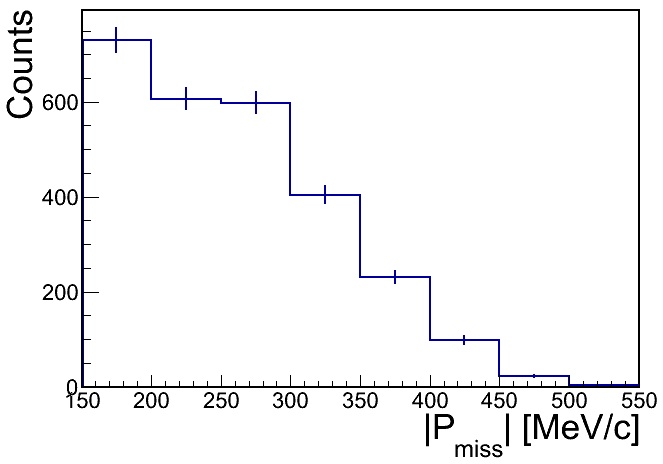}
\caption[]{\label{fig:Pmiss1D} \baselineskip 13pt (left) The expected number of events
as a function of missing momentum for the  kinematic points of
Table~\ref{tab:kin}; (right) the number of events for all
kinematic points combined, starting at $p_m = 150$ MeV/c.}
\end{center}
\end{figure}

\begin{figure}[htpb]
\begin{center}
\includegraphics[width=3in]{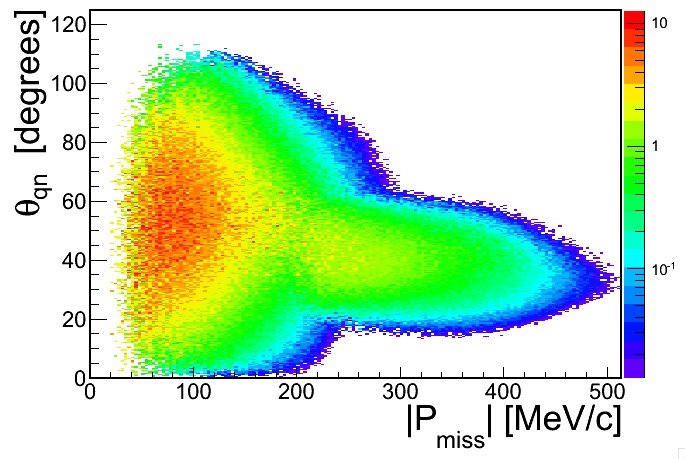}
\includegraphics[width=3in]{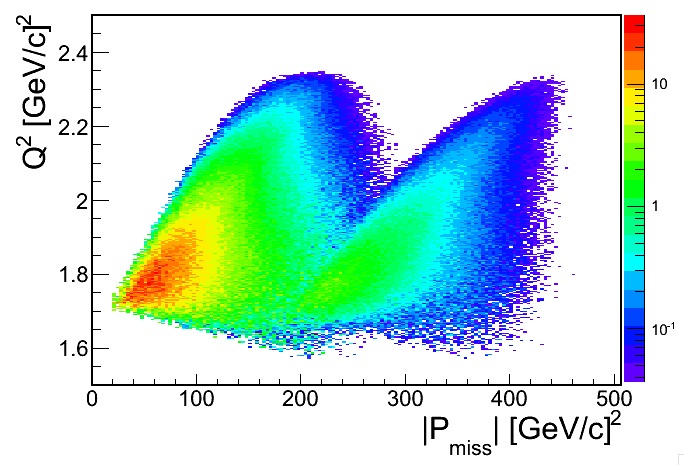}
\caption[]{\label{fig:ThetaQnPmiss} \baselineskip 13pt (left) $\theta_{rq}$, the angle
  between the recoiling system (missing momentum) and $\vec q$, versus
  the missing momentum.  
%The dashed line shows the cut at  $\theta_{rq}\le 40^\circ$. 
(right) $Q^2$, the square of the momentum
  transfer plotted versus the missing momentum.  The vertical scales are
logarithmic.}
\end{center}
\end{figure}

\begin{figure}[htpb]
\begin{center}
\includegraphics[width=6in]{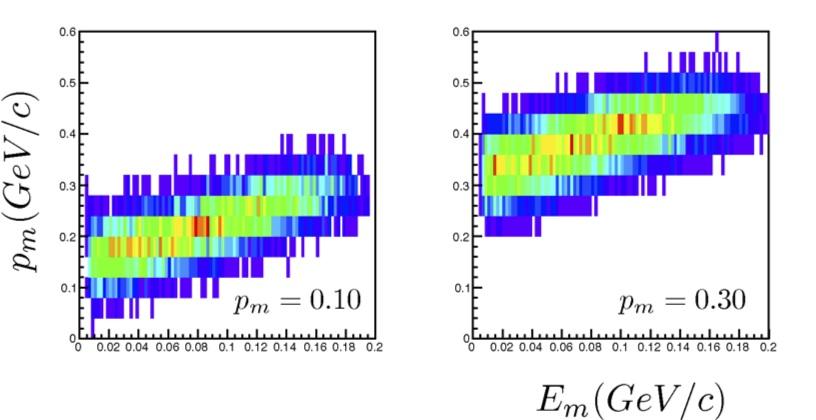}
\end{center}
\caption[]{\label{fig:EmissPmiss} \baselineskip 13pt The missing
  energy -- missing momentum distribution for the two
  kinematic settings.  This shows the kinematic coverage and does not
  include cross section weighting.}
\end{figure}

The yields as a function of missing momentum are shown in Fig.{}
\ref{fig:Pmiss1D} for the three kinematic points.
Figs. \ref{fig:ThetaQnPmiss} and \ref{fig:EmissPmiss} show the
distribution of events.

\subsection{Expected Results}

\begin{figure}[htpb]
\begin{center}
\includegraphics[width=5in]{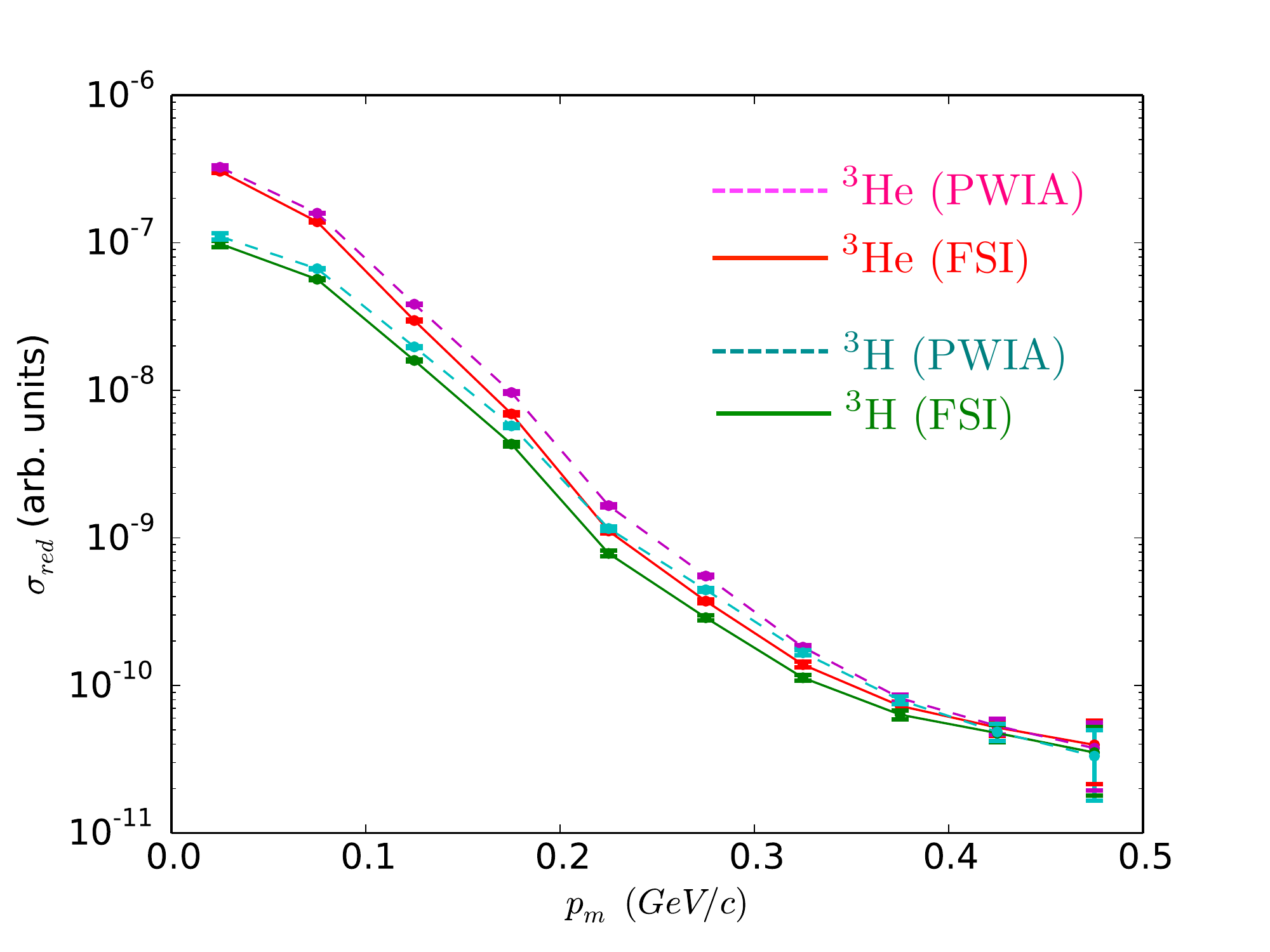}
\end{center}
\caption[]{\label{fig:CrossSectionErrors} \baselineskip 13pt The expected
  statistical uncertainties in the proposed measured reduced cross
  sections for
  \het\eep{} and \trit\eep{} integrated over missing energy as a function of
  missing momentum.  The results are shown for calculations both
  without (PWIA) and with (FSI) the effects of final state
  interactions \cite{misakpc}.
  }
\end{figure}

\begin{figure}[htpb]
\begin{center}
\includegraphics[width=5in]{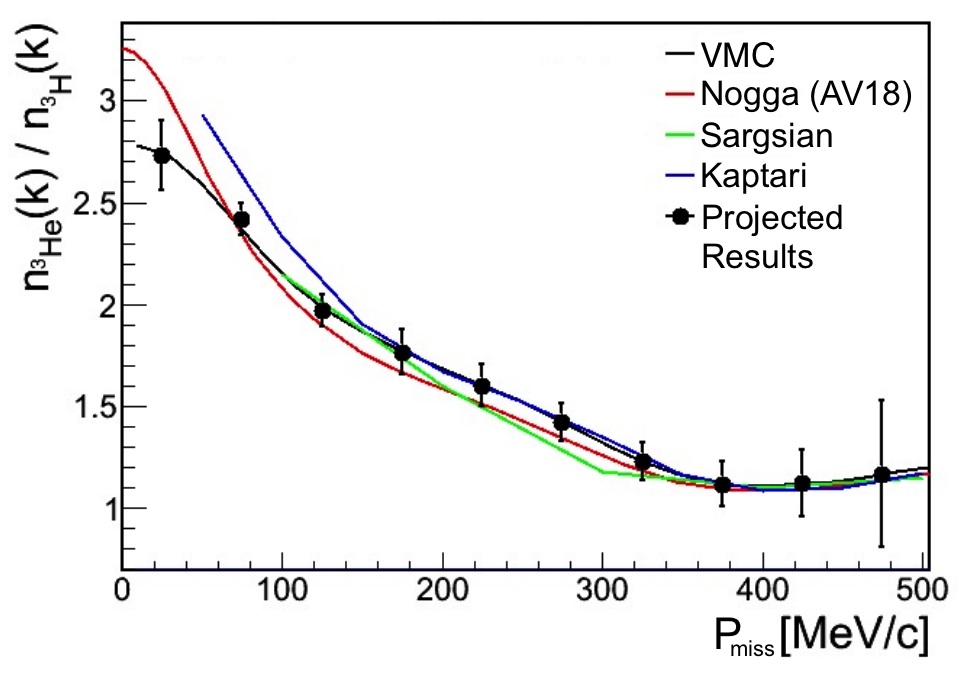}
\end{center}
\caption[]{\label{fig:A3Ratio} \baselineskip 13pt The expected
  statistical uncertainties in the proposed measured cross section ratios of
  \het\eep{}/\trit\eep{} as a function of
  missing momentum.  
  }
\end{figure}

\begin{figure}[htpb]
\begin{center}
\includegraphics[width=5in]{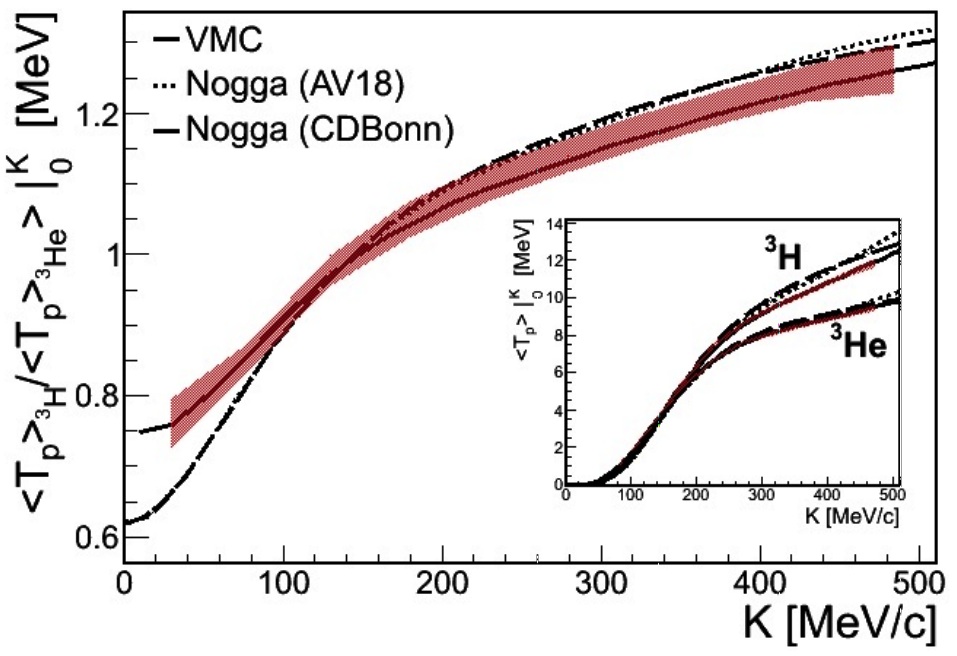}
\end{center}
\caption[]{\label{fig:A3TpRatio} \baselineskip 13pt The expected
  statistical uncertainties in the  ratios of the average
  nucleon kinetic energies
 measured in  \het\eep{} and \trit\eep{} as a function of
  missing momentum (see Eq. \ref{eq:Tp}).  The red band shows the
  total expected uncertainty of the experimental integration.
  }
\end{figure}

We will measure the  \het\eep{} and \trit\eep{} cross sections
at each value of missing momentum
by integrating the cross sections over missing energy.  We will then
construct the cross section ratio \het\eep/\trit\eep.  The expected
statistical uncertainties are shown in Figs.{}
\ref{fig:CrossSectionErrors} and \ref{fig:A3Ratio}.  The expected
results for the ratio of kinetic energies is shown in Fig.{} \ref{fig:A3TpRatio}.

\section{Summary}

Fully understanding few-nucleon systems is vital to our understanding
of nuclear physics.  We propose to take advantage of a unique
opportunity to measure the cross sections of \trit\eep{} and
\het\eep{} in Hall A in order to determine the momentum distributions
of the minority and majority fermions in asymmetric nuclei.  \trit\eep{} has never been measured before and
\het\eep{} has never been measured in kinematics which minimize the
effects of final state interactions (FSI), meson exchange currents,
and isobar configurations.  

By forming the cross section ratio of \het\eep{} to \trit\eep, the
remaining effects of FSI almost entirely cancel, allowing us to
extract the ratio of their momentum distributions with unprecedented
precision.  We propose to measure this ratio from $p_m = 0$ where
independent nucleons dominate (where the effect of SRC increases the
ratio from the naive expectation due to proton counting of two) to
$p_m = 0.5$ GeV/c where nucleons belonging to short range correlations
(SRC) dominate (and the naive ratio due to pair counting should be
about one).

We also propose to measure the absolute cross sections of these reactions in
order to provide a stringent test of theoretical models.  These models
will face different challenges in calculating the absolute cross
sections and the cross section ratio \het\eep/\trit\eep.

This will allow us to measure the majority and minority fermion
momentum distributions in the most asymmetric stable nuclei.  This
will be a strong test of modern calculations in asymmetric nuclei.

This proposal was deferred by PAC 40.  In this  resubmission, we have
reduced the requested beam time from 30 to 10 days by focussing on
\het{} and \trit, omitting the deuterium target and reducing
the beam time at large missing momentum.  We also emphasize the
importance of measuring majority and minority fermion momentum
distributions in asymmetric nuclei.

We propose to measure these cross sections using the MARATHON target
and the Hall A HRS spectrometers in their standard configuration.
Radioactive targets in general and tritium targets in particular pose
serious safety issues and are thus very difficult to install.  The
MARATHON target will only be at Jefferson Lab for a brief period.
This will be the only opportunity to measure \trit\eep.  It is crucial
to take advantage of this opportunity to fill a gaping hole in our
knowledge of few-nucleon systems.

We request 10 days of beam time to measure \trit\eep{} and \het\eep.

\eject
\bibliography{emc,eep,3he}

\end{document}